\documentclass[11pt]{article}
\usepackage{amsmath,amsfonts,amssymb}
\numberwithin{equation}{section}

\newcommand{\nc}{\newcommand}
\nc{\bib}{\bibitem}
\nc{\al}{\alpha}
\nc{\g}{\gamma}
\nc{\G}{\Gamma}
\nc{\D}{\Delta}
\nc{\eps}{\epsilon}
\nc{\vareps}{\varepsilon}
\nc{\la}{\lambda}
\nc{\La}{\Lambda}
\nc{\var}{\varphi}
\nc{\pa}{\partial}
\nc{\nn}{\nonumber \\ }
\nc{\be}{\begin{equation}}
\nc{\ee}{\end{equation}}
\nc{\bea}{\begin{eqnarray}}
\nc{\eea}{\end{eqnarray}}
\nc{\bra}[1]{\langle {#1}|}
\nc{\ket}[1]{|{#1}\rangle}
\nc{\gb}{\bar{g}}
\nc{\sbar}{\bar{s}}
\nc{\Ab}{\bar{A}}
\nc{\Db}{\bar{D}}
\nc{\Lc}{\mathcal{L}}
\nc{\Oc}{\mathcal{O}}
\nc{\Qh}{\hat{Q}}
\nc{\hh}{\hat{h}}
\nc{\kh}{\hat{k}}
\nc{\Kc}{\mathcal{K}}

\begin{document}

\topmargin -5mm
\oddsidemargin 5mm

\setcounter{page}{1}

\vspace{8mm}
\begin{center}
{\huge {\bf A near-NHEK/CFT correspondence}}

\vspace{8mm}
 {\LARGE J{\o}rgen Rasmussen}
\\[.3cm]
 {\em Department of Mathematics and Statistics, University of Melbourne}\\
 {\em Parkville, Victoria 3010, Australia}
\\[.4cm]
 {\tt j.rasmussen@ms.unimelb.edu.au}

\end{center}

\vspace{8mm}
\centerline{{\bf{Abstract}}}
\vskip.4cm
\noindent
We consider excitations around the recently introduced near-NHEK metric
describing the near-horizon geometry of the near-extremal four-dimensional Kerr black hole.
This geometry has a $U(1)_L\times U(1)_R$ isometry group which can be enhanced
to a pair of commuting Virasoro algebras. We present boundary conditions for which
the conserved charges of the corresponding asymptotic symmetries are well defined and
non-vanishing and find the central charges $c_L=12J/\hbar$ and $c_R=0$ where 
$J$ is the angular momentum of the black hole.
Applying the Cardy formula reproduces the
Bekenstein-Hawking entropy of the black hole. This suggests that the near-extremal
Kerr black hole is holographically dual to a non-chiral two-dimensional
conformal field theory.
\renewcommand{\thefootnote}{\arabic{footnote}}
\setcounter{footnote}{0}

\newpage

\section{Introduction}

The near-horizon geometry of an extremal four-dimensional
Kerr black hole \cite{BH9905} is described by the so-called
NHEK metric. According to the recently conjectured Kerr/CFT correspondence \cite{GHSS0809},
the corresponding quantum theory is holographically dual to a chiral 
Conformal Field Theory (CFT) in two dimensions. 
In the spirit of~\cite{BH86} and using the formalism of~\cite{BB0111}, 
it was found that certain boundary conditions enhance
the $U(1)_L$ symmetry of the $U(1)_L\times SL(2,\mathbb{R})_R$ isometry group to
a Virasoro algebra. Strong evidence \cite{GHSS0809} for the correspondence is found in the exact
agreement between the macroscopic Bekenstein-Hawking entropy \cite{Bek73}
of the black hole and the Cardy formula for the CFT entropy.
This analysis has been successfully generalized and applied to a variety of extremal
black holes \cite{wake1,wake1a,wake2,wake2a,wake3,AHMR0906,wake4,wake5}.
Extending the Kerr/CFT correspondence to the {\em near}-extremal Kerr black hole, however,
has presented some serious challenges. The dual two-dimensional CFT should be non-chiral. 
It is the main objective of the present work 
to offer a possible resolution to one of these obstacles. 

Boundary conditions enhancing the 
$SL(2,\mathbb{R})_R$ isometries to a Virasoro algebra were examined in
\cite{MTY0907,Ras0908}.
Since no self-consistent set of boundary conditions was found which enhances 
the $U(1)_L$ isometry at the same time, these
works do not capture the full non-chiral CFT in the dual picture. 
Several results on near-extremal black holes have since been obtained \cite{CaLa0908}
and have provided significant insight.
Extending the approach of \cite{MS9609}, 
the work \cite{BHSS0907} on black-hole superradiance and the subsequent generalizations thereof
\cite{CL0908} have provided further and highly non-trivial
support for the Kerr/CFT correspondence.
These results on the near-extremal Kerr black hole are all based on deviations from the
geometrical approach of \cite{GHSS0809} since consistent boundary conditions
which allow for both left- and right-moving sectors have not yet been identified.
Another very recent development is the analysis of a so-called hidden conformal 
symmetry \cite{CMS1004} which is not a symmetry of the spacetime geometry.

Here we reapply the original approach of \cite{GHSS0809} to the near-extremal Kerr
black hole. Its near-horizon geometry is described by the so-called near-NHEK metric
derived in \cite{BHSS0907}.
This geometry has a $U(1)_L\times U(1)_R$ isometry group which we find can be enhanced
to a pair of commuting Virasoro algebras. We present the corresponding 
boundary conditions and argue that the conserved charges of the corresponding asymptotic 
symmetries are well defined and non-vanishing. 
The central charges are $c_L=12J/\hbar$ and $c_R=0$ where 
$J$ is the angular momentum of the black hole.
Applying the Cardy formula reproduces the
Bekenstein-Hawking entropy of the black hole. This supports the assertion that the near-extremal
Kerr black hole is holographically dual to a {\em non-chiral} two-dimensional CFT.
We refer to the corresponding duality as the near-NHEK/CFT correspondence since the
original Kerr/CFT correspondence \cite{GHSS0809} actually concerns a duality of 
the form NHEK/CFT.

The Virasoro algebra has been observed in other black-hole contexts as well.
Extending the work~\cite{Str9712} on the entropy of three-dimensional black holes such as the
BTZ black hole~\cite{BTZ9204}, it was found in~\cite{Car9812,Sol9812} 
that a copy of the Virasoro algebra appears in the near-horizon region of any black hole
and that it reproduces the black-hole entropy using the Cardy formula.
Free-field approaches to the dual description of black holes 
are discussed in \cite{Sol9812,Ras0909}, while
black-hole solutions of gravity theories
with higher-derivative corrections are considered
in~\cite{SS9909} and references therein.

\section{Kerr/CFT correspondence}
\label{SecKerr}

\subsection{NHEK geometry}

The Near-Horizon
Extremal Kerr (NHEK) geometry \cite{BH9905,GHSS0809} is described by
\be
 d\bar{s}^2=2GJ\Gamma\Big(\!-r^2dt^2+\frac{dr^2}{r^2}+d\theta^2
  +\Lambda^2\big(d\phi+rdt\big)^2\Big)
\label{ds2Kerr}
\ee
where
\be
 \Gamma=\Gamma(\theta)=\frac{1+\cos^2\theta}{2},\qquad\qquad
  \La=\Lambda(\theta)=\frac{2\sin\theta}{1+\cos^2\theta}
\label{GaLa}
\ee
and where $\theta\in[0,\pi]$ and $\phi\sim\phi+2\pi$.
The corresponding isometry group $U(1)_L\times SL(2,\mathbb{R})_R$ is generated by
\be
 \big\{\pa_\phi\big\}\cup
   \big\{\pa_t,t\pa_t-r\pa_r,\big(t^2+\frac{1}{r^2}\big)\pa_t-2tr\pa_r-\frac{2}{r}\pa_\phi\big\}
\label{isoKerr}
\ee
while the ADM mass $M$ and angular momentum $J$ of the associated extremal
Kerr black hole are related through
\be
 J=GM^2
\ee

\subsection{Conserved charges}

The interest here is in perturbations $h_{\mu\nu}$
of the near-horizon geometry of the extremal black hole
whose background metric $\gb_{\mu\nu}$ is defined in (\ref{ds2Kerr}).
Asymptotic symmetries are generated by the diffeomorphisms whose action on the
metric generates metric fluctuations compatible with the chosen boundary conditions.
We are thus looking for contravariant vector fields $\eta$ along which the Lie derivative
of the metric is of the form 
\be
 \Lc_\eta \gb_{\mu\nu}\sim h_{\mu\nu}
\label{Lgh}
\ee
To such an asymptotic symmetry generator $\eta$, one associates~\cite{BB0111}
the conserved charge
\be
 Q_\eta=\frac{1}{8\pi G}\int_{\pa\Sigma}\sqrt{-\gb}k_\eta[h;\gb]
   =\frac{1}{8\pi G}\int_{\pa\Sigma}\frac{\sqrt{-\gb}}{4}\eps_{\al\beta\mu\nu}d_\eta^{\mu\nu}[h;\gb]
     dx^\al\wedge dx^\beta
\label{Q}
\ee
where
\be
 d_\eta^{\mu\nu}[h;\gb]
  =\eta^\nu \Db^\mu h-\eta^\nu \Db_\sigma h^{\mu\sigma} +\eta_\sigma\Db^\nu h^{\mu\sigma}
  -h^{\nu\sigma}\Db_\sigma\eta^\mu+\frac{1}{2}h\Db^\nu\eta^\mu
  +\frac{1}{2}h^{\sigma\nu}\big(\Db^\mu\eta_\sigma+\Db_\sigma\eta^\mu\big)
\label{dmunu}
\ee
and where $\pa\Sigma$ is the boundary of a three-dimensional
spatial volume, ultimately near spatial infinity. Here, indices are lowered and raised
using the background metric $\gb_{\mu\nu}$ and its inverse, 
$\Db_\mu$ denotes a background covariant derivative, 
while $h$ is defined as $h=\gb^{\mu\nu}h_{\mu\nu}$.
To be a well-defined charge in the asymptotic limit, 
the underlying integral must be finite as $r\to\infty$.
If the charge vanishes, the asymptotic symmetry is rendered trivial.
The asymptotic symmetry group is generated by the diffeomorphisms 
whose charges are well-defined and non-vanishing.
The algebra generated by the set of well-defined charges is governed
by the Dirac brackets computed~\cite{BB0111} as
\be
 \big\{Q_\eta,Q_{\hat\eta}\big\}
  =Q_{[\eta,\hat\eta]}+\frac{1}{8\pi G}\int_{\pa\Sigma}\sqrt{-\gb}k_{\eta}[\Lc_{\hat\eta}\gb;\gb]
\label{QQ}
\ee
where the integral yields the eventual central extension.

\subsection{Boundary conditions}

Written in the ordered basis $\{t,r,\phi,\theta\}$, the boundary conditions considered
in~\cite{GHSS0809} are the fall-off conditions
\be
 h_{\mu\nu}=\Oc\!\left(\!\!\begin{array}{cccc} r^2&r^{-2}&1&r^{-1} \\ &r^{-3}&r^{-1}&r^{-2} \\ 
   &&1&r^{-1} \\ &&&r^{-1} \end{array} \!\!\right),
  \qquad\quad h_{\mu\nu}=h_{\nu\mu}
\label{hKerr}
\ee
and the zero-energy condition 
\be
 Q_{\pa_t}=0
\label{zero}
\ee 
Consistency of these conditions was confirmed in~\cite{AHMR0906}.
The generators of the corresponding asymptotic symmetry group read
\be
 \xi=-\eps'(\phi)r\pa_r+\eps(\phi)\pa_{\phi}
\label{xi}
\ee 
and form the centreless Virasoro algebra 
\be
 \big[\xi_{\eps},\xi_{\hat\eps}\big]=\xi_{\eps\hat\eps'-\eps'\hat\eps}
\label{Virdiff}
\ee
This symmetry is an enhancement of the exact $U(1)_L$ isometry generated by the Killing vector 
$\pa_\phi$ of (\ref{ds2Kerr}) as the latter is recovered by setting $\eps(\phi)=1$. 
The usual form of the Virasoro algebra 
is obtained by choosing an appropriate basis for the functions $\eps(\phi)$ and $\hat\eps(\phi)$,
where we recall the periodicity $\phi\sim\phi+2\pi$.
With respect to the basis $\xi_n(\phi)$, where 
\be
 \eps_n(\phi)=-e^{-in\phi}
\label{eps}
\ee
one introduces the dimensionless quantum versions 
\be
 L_n=\frac{1}{\hbar}\Big(Q_{\xi_n}+\frac{3J}{2}\delta_{n,0}\Big)
\label{L}
\ee
of the conserved charges. After the usual substitution $\{.,.\}\to-\frac{i}{\hbar}[.,.]$ of 
Dirac brackets by quantum commutators, the quantum charge algebra is 
recognized~\cite{GHSS0809} as the centrally-extended Virasoro algebra
\be
 \big[L_n,L_m\big]=(n-m)L_{n+m}+\frac{c}{12}n(n^2-1)\delta_{n+m,0},\qquad\quad
   c_L=\frac{12J}{\hbar}
\label{VirKerr}
\ee
This quantum charge algebra also arises when considering boundary conditions
sufficiently similar to those in (\ref{hKerr}).
A partial classification of such alternatives can be found in~\cite{Ras0908}.

\section{Near-NHEK/CFT correspondence}
\label{SecNearNHEK}

\subsection{Near-NHEK geometry}

We are interested in infinitesimal excitations above extremality of the Kerr black hole.
To describe this near-extremal Kerr black hole, we follow \cite{BHSS0907}
and consider a generalization of the NHEK geometry in which the temperature of the
near-horizon geometry is fixed and non-zero. This temperature is denoted by $T_R$
and the corresponding near-NHEK geometry is described by
\be
 d\sbar^2=2J\Gamma\Big(\!-r(r+2\al)dt^2+\frac{dr^2}{r(r+2\al)}+d\theta^2
  +\Lambda^2\big(d\phi+(r+\al)dt\big)^2\Big)
\label{ds2nearNHEK}
\ee
where
\be
 \al=2\pi T_R
\ee
while $\Gamma$ and $\La$ are given in (\ref{GaLa}). Here and in the following, we
use the unit convention of \cite{BHSS0907} where $G=\hbar=c=1$.
We denote the background metric (\ref{ds2nearNHEK}) by $\gb$, as we
did with the NHEK geometry (\ref{ds2Kerr}), and hope that no confusion will
arise from this abuse of notation.
The NHEK geometry (\ref{ds2Kerr}) follows immediately from the near-NHEK geometry
(\ref{ds2nearNHEK}) by setting $T_R=0$.
We also note that the determinant of the near-NHEK metric $\gb$ only depends on $\theta$ as
\be
 \sqrt{-\gb}=4J^2\Gamma^2(\theta)\La(\theta)=2J^2\!\sin\theta(1+\cos^2\theta)
\ee

It is readily verified that
\be
 \big\{\pa_\phi\big\}\cup\big\{\pa_t\big\}
\ee
generate exact $U(1)_L\times U(1)_R$ isometries of the near-NHEK geometry.
The $U(1)_R$ subgroup of the $SL(2,\mathbb{R})_R$ isometries of the NHEK geometry 
is thus generated by the lowering (or raising) operator
of $SL(2,\mathbb{R})_R$, not by the Cartan generator of the corresponding Lie algebra.

\subsection{Boundary conditions and conserved charges}

First, we modify the fall-off conditions (\ref{hKerr}) by introducing 
\be
 \hh_{\mu\nu}=\Oc\!\left(\!\!\begin{array}{cccc} 
   r^2&r^{-3}&r^{1}&r^{-2} \\ 
   &r^{-4}&r^{-1}&r^{-3} \\ 
   &&1&r^{-2} \\ 
   &&&r^{-2} \end{array} \!\!\right),
  \qquad\quad \hh_{\mu\nu}=\hh_{\nu\mu}
\label{hnearNHEK}
\ee
and find the asymptotic Killing vectors
\bea
 K_\eps&=&\big[\Oc(r^{-4})\big]\pa_t+\big[-(r+\al)\eps'(\phi)+\Oc(r^{-1})\big]\pa_r
   +\big[\eps(\phi)+\Oc(r^{-3})\big]\pa_\phi
   +\big[\Oc(r^{-2})\big]\pa_\theta   \nn
 \Kc_\vareps&=&\big[\vareps(t)+\Oc(r^{-4})\big]\pa_t+\big[\Oc(r^{-1})\big]\pa_r
   +\big[\Oc(r^{-3})\big]\pa_\phi
   +\big[\Oc(r^{-2})\big]\pa_\theta
\eea
where $\eps(\phi)$ and $\vareps(t)$ are smooth functions.
The generators of the corresponding asymptotic symmetries read
\be
 \xi=-(r+\al)\eps'(\phi)\pa_r+\eps(\phi)\pa_\phi,\qquad
 \zeta=\vareps(t)\pa_t
\label{xizeta}
\ee
and form a commuting pair of centreless Virasoro algebras
\be
 \big[\xi_\eps,\xi_{\hat{\eps}}\big]=\xi_{\eps\hat{\eps}'-\eps'\hat{\eps}},\qquad
 \big[\zeta_\vareps,\zeta_{\hat{\vareps}}\big]=\zeta_{\vareps\hat{\vareps}'-\vareps'\hat{\vareps}},
     \qquad
 \big[\xi_\eps,\zeta_\vareps\big]=0
\ee
Along $\xi$ and $\zeta$, the Lie derivatives of the near-NHEK metric are worked out to be
\bea
 \mathcal{L}_\xi\gb_{\mu\nu}&=&2J\Gamma\left(\begin{array}{cccc}
  -2(\La^2-1)(r+\al)^2\eps'(\phi)&0&0&0\\[.2cm]
  0&\frac{2\al^2\eps'(\phi)}{r^2(r+2\al)^2}&-\frac{(r+\al)\eps''(\phi)}{r(r+2\al)}&0\\[.2cm]
  0&-\frac{(r+\al)\eps''(\phi)}{r(r+2\al)}&2\La^2\eps'(\phi)&0\\[.2cm]
  0&0&0&0
  \end{array}\right)\nn
 \mathcal{L}_\zeta\gb_{\mu\nu}&=&2J\Gamma\vareps'(t)\left(\begin{array}{cccc}
  2(\La^2-1)r(r+2\al)+2\al^2\La^2&0&\La^2(r+\al)&0\\[.2cm]
  0&0&0&0\\[.2cm]
  \La^2(r+\al)&0&0&0\\[.2cm]
  0&0&0&0
  \end{array}\right)
\label{Lie}
\eea

Second, we supplement the fall-off conditions (\ref{hnearNHEK})
by a weakened zero-energy condition (\ref{zero}) where instead of requiring
$Q_{\pa_t}=0$, we allow this conserved charge to be proportional to $\al$.
An inspection of the asymptotic $r$-expansion of $k_{\pa_t}[\hh;\gb]$ reveals that the
divergent term (linear in $r$) is independent of $\al$ while the constant term 
(independent of $r$) is at most linear in $\al$. Subleading terms in $r$ can here be ignored
as $r\to\infty$. We thus impose the condition
\be
 \big(1-\al\pa_\al\big)Q_{\pa_t}=0
\label{aQ0}
\ee
Under this constraint, only perturbations $\hh$ preserving it and only background metrics
$g$ which can be reached from the near-NHEK geometry via a path of such perturbations are
considered. As in the similar situation \cite{GHSS0809} 
arising when imposing (\ref{zero}) in addition to (\ref{hKerr}),
this is presumably a complicated nonlinear submanifold of the geometries allowed
by the linear boundary conditions (\ref{hnearNHEK}).

Third, $Q_{\zeta}$ must be well-defined for all $\zeta$, not just for $\vareps(t)=1$ 
as in (\ref{aQ0}), so we should study the consequences of the constraint (\ref{aQ0})
on the asymptotic expansion of $Q_{\zeta}$. To this end, we examine the 
integrands of the corresponding charges (\ref{Q}) and obtain the remarkably simple 
expression
\be
 \sqrt{-\gb}\big(k_\zeta[\hh;\gb]-\vareps(t)k_{\pa_t}[\hh;\gb]\big)\big|_{d\phi\wedge d\theta}=
  -\frac{1}{4}\La\vareps'(t)(r+\al)\hh_{r\phi}
\label{kk}
\ee
valid for all $r$.
It follows, in particular, that the divergent part of the asymptotic $r$-expansion
of $k_\zeta[\hh;\gb]$
matches the divergent part of $\vareps(t)k_{\pa_t}[\hh;\gb]$ where
\bea
  &&\sqrt{-\gb}k_{\pa_t}[\hh;\gb]\big|_{d\phi\wedge d\theta}=
   \frac{1}{4\La}\Big(-\La^4r^{-1}\hh_{tt}+2\La^2(\La^2-r\pa_r)\hh_{t\phi}
   +2(1-\La^2)r^2\pa_\phi\hh_{r\phi}\nn
  &&\hspace{7.5cm}-(\La^4+\La^2-2)r\hh_{\phi\phi}\Big)+\Oc(r^0)
\label{div}
\eea 
Here we have used that $\pa_r\hh_{\phi\phi}=\Oc(r^{-2})$.
This matching ensures that the conserved charge $Q_{\zeta}$ is {\em finite} 
when (\ref{hnearNHEK}) and (\ref{aQ0}) are satisfied.

Fourth, and despite the results above, the expression (\ref{kk}) actually indicates that 
(\ref{hnearNHEK}) and (\ref{aQ0}) provide an inconsistent set of boundary conditions.
The problem is that, while contravariant vector fields of the form $\zeta$ in (\ref{xizeta})
satisfy the Jacobi identity for commutators, 
the corresponding conserved charges $Q_\zeta$ may not satisfy the Jacobi identity for Dirac
brackets. This is illustrated by $\vareps_j(t)=t^{n_j}$, $j=1,2,3$, if 
$n_2=1-n_1$ with $n_1$ and $n_3$ generic, since then
$Q_{[\zeta_1,\zeta_2]}=(1-2n_1)Q_{\pa_t}=0$.
We should therefore prevent the term (\ref{kk}) from contributing to
the surface integral (\ref{Q}) in the definition of $Q_\zeta$.
However, $\mathcal{L}_\xi\gb_{r\phi}$ goes like $r^{-1}$ asymptotically so we cannot
strengthen the fall-off condition $\hh_{r\phi}$ without affecting the conserved charge $Q_\xi$.
As a remedy, one may require that $\hh_{r\phi}$ is a total $\phi$-derivative 
on the boundary. Such a requirement will presumably also reduce possible 
back-reaction effects, see \cite{AHMR0906} on the extremal Kerr/CFT correspondence. 
We therefore suggest to impose the condition
\be
 \hh_{r\phi}\big|_{\pa\Sigma}=\pa_\phi\hat{H}_{r\phi}
\label{hH}
\ee
and note that it is satisfied by the perturbations (\ref{Lie}) generated by $\xi$ and $\zeta$.
The combined set of boundary conditions given in (\ref{hnearNHEK}), (\ref{aQ0}) 
and (\ref{hH}) should then ensure that $Q_\xi$ and $Q_\zeta$ are {\em well defined}.

Now, taking the constraint (\ref{aQ0}) explicitly into account, $Q_{\zeta}$ can be written as
\be
 Q_\zeta=\frac{1}{8\pi}\int_{\pa\Sigma}\sqrt{-\gb}\kh_\zeta[\hh;\gb]
\label{Qzeta}
\ee
where
\be
 \sqrt{-\gb}\kh_\zeta[\hh;\gb]
   =\frac{\al\vareps(t)}{4\La}\Big(\La^4r^{-2}\hh_{tt}-2\La^2\pa_r\hh_{t\phi}
     -(\La^4+\La^2-2)\hh_{\phi\phi}\Big)d\phi\wedge d\theta+\ldots
\label{kh}
\ee
The dots indicate that terms not contributing to the charge (\ref{Qzeta}) have been omitted.
In particular, total $\phi$-derivatives can be ignored. 
We also find that the integrand in
(\ref{Q}) for the conserved charge $Q_\xi$ is given by
\bea
 &&\sqrt{-\gb}k_\xi[\hh;\gb]= \frac{\La}{2}\Big(
   \eps(\phi)\big[
 -\frac{\La^2\hh_{tt}}{2r^2}
 +\frac{(\La^2-r\pa_r)\hh_{t\phi}}{r}
 -r\pa_\phi\hh_{r\phi}
  -\frac{(\La^2+1)\hh_{\phi\phi}}{2}
 \big]\nn
 &&\hspace{3.2cm}
 +\eps'(\phi)\big[
  r\hh_{r\phi}
  -\frac{\pa_\phi\hh_{t\phi}}{\La^2r}\big]
 +\eps''(\phi)\big[\frac{\hh_{t\phi}}{2\La^2r}\big]
 \Big)d\phi\wedge d\theta+\ldots
\eea

We finally argue that the conserved charge $Q_\zeta$ is {\em non-vanishing}.
Separating the boundary condition
(\ref{aQ0}) into a vanishing condition on the divergent part of $Q_{\pa_t}$ 
(computed using (\ref{div})) and a vanishing condition on the non-divergent part could render 
$Q_\zeta$ trivial. We find that this is not the case for general perturbations satisfying
the fall-off conditions (\ref{hnearNHEK}). Instead, we find that a simple 
re-scaling of the divergent part by a factor of $\al/r$ 
can be used to simplify the integrand of (\ref{Qzeta})
\be
 \sqrt{-\gb}\kh_\zeta[\hh;\gb]
   \to\frac{\al\La^3\vareps(t)}{2}\Big(\frac{\hh_{tt}}{r^2}-\frac{\hh_{t\phi}}{r}
   \Big)d\phi\wedge d\theta+\ldots
\label{kh2}
\ee
It is recalled that the NHEK geometry is obtained by setting $\al=0$ in which
case the conserved charge $Q_\zeta$ is seen to vanish.
The conserved charge $Q_\xi$, on the other hand, is unaffected by setting $\al=0$.

\subsection{Central charges}

First, we note that the left-moving sector obtained from the conserved charges
$Q_\xi$ is generated by the same Virasoro modes (\ref{L}) forming the same
Virasoro algebra (\ref{VirKerr}) as in the NHEK case. The corresponding central charge is thus
given by
\be
 c_L=\frac{12J}{\hbar}
\ee

To obtain a quantum algebra in the right-moving sector, one can perform an analytic 
continuation of $t$ and introduce \cite{MTY0907}
\be
 L_n=\oint\frac{dt}{2\pi i}Q_{\zeta_n},\qquad \zeta_n=\vareps_n(t)\pa_t,\qquad 
     \vareps_n(t)=-t^{n+1}
\ee
Since $\sqrt{-\gb}k_\zeta[\mathcal{L}_{\hat{\zeta}}\gb;\gb]$ is linear in $\vareps(t)\hat{\vareps}'(t)$
and therefore only involves a single derivative, the eventual central extension of the 
corresponding Virasoro algebra can be absorbed in a redefinition of the quantum 
generator $L_{-2}$, much akin to the redefinition of $L_0$ in (\ref{L}).
This implies that the central charge of the right-moving sector vanishes
\be
 c_R=0
\label{cR}
\ee

By evaluating $\sqrt{-\gb}k_\xi[\mathcal{L}_\zeta\gb;\gb]$ and 
$\sqrt{-\gb}k_\zeta[\mathcal{L}_\xi\gb;\gb]$ explicitly, it is verified that there is no central charge
mixing the two conformal sectors. This establishes that the two Virasoro algebras are
indeed mutually commutative.

\subsection{Entropy}

The quantum theory in the Frolov-Thorne vacuum \cite{FT89} restricted to the extremal
Kerr black hole has the left-moving temperature \cite{GHSS0809}
\be
 T_L=\frac{1}{2\pi}
\label{TL}
\ee
Applying the Cardy formula to the dual CFT yields the entropy
\be
 S=\frac{\pi^2}{3}\big(c_L T_L+c_R T_R\big)
\label{S}
\ee
For NHEK, the dual CFT has $c_L=\tfrac{12J}{\hbar}$ and is chiral (implying $c_R=T_R=0$) 
thereby reproducing the Bekenstein-Hawking entropy \cite{Bek73}
\be
 S_{BH}=\frac{2\pi J}{\hbar}
\label{SBH}
\ee
In the case of the near-extremal Kerr black hole, the near-horizon geometry is described by the 
near-NHEK metric. By construction, the right-moving sector of the dual {\em non-chiral} 
CFT is excited with the non-zero temperature $T_R=\tfrac{\al}{2\pi}$ 
while we have found that the corresponding
central charge $c_R$ vanishes (\ref{cR}).
The left-moving sector is characterized by the same $c_L$ and $T_L$ as in the extremal case.
The entropy is therefore given by the same expression (\ref{SBH}) as in the extremal case.

\subsection*{Acknowledgments}
\vskip.1cm
\noindent
This work is supported by the Australian Research Council. 
The author thanks Omar Foda for discussions.


\end{document}